\documentclass{webofc}
\usepackage[varg]{txfonts}   
\usepackage{float}
\usepackage[colorlinks=true, allcolors=black]{hyperref}

\newcommand{\trecento}{\textsc{The Three Hundred} }

\begin{document}
\title{Impact of filaments on galaxy cluster properties in \textrm{\textmd{\textsc{The Three Hundred}}} simulation}
%
%

\author{\lastname{S. Santoni}\inst{1}\fnsep\thanks{\email{sara.santoni@uniroma1.it}} \and
        \lastname{M. De Petris}\inst{1} \and
        \lastname{A. Ferragamo}\inst{1} \and
        \lastname{G. Yepes}\inst{2} \and
        \lastname{W. Cui}\inst{2,3}
}

\institute{Dipartimento di Fisica, Sapienza Università di Roma, Piazzale Aldo Moro 5, I-00185 Rome, Italy
\and
Departamento de Fısica Teorica \& CIAFF, Facultad de Ciencias, Universidad Autonoma de Madrid, Modulo 8, E-28049 Madrid, Spain
 \and
 Institute for Astronomy, University of Edinburgh, Edinburgh EH9 3HJ, UK}
\abstract{%
  Galaxy clusters and their filamentary outskirts reveal useful laboratories to test cosmological models and investigate Universe composition and evolution. Their environment, in particular the filaments of the Cosmic Web to which they are connected, plays an important role in shaping the properties of galaxy clusters. In this project, we analyse the gas filamentary structures present in 324 regions of \trecento hydrodynamical simulation extracted with the DisPerSE filament finder. We estimate the number of gas filaments globally connected to several galaxy clusters, i.e. the connectivity \textit{k}, with a mass range of $10^{13} \leq  M_{200} \, h^{-1} \, M_{\odot} \leq 10^{15} $ at redshift $z=0$. We study the positive correlation between the connectivity and mass of galaxy clusters. Moreover, we explore the impact of filaments on the dynamical state of clusters, quantified by the degree of relaxation parameter $\chi$. 
}
\maketitle
\section{Introduction}
\label{intro}
Clusters of galaxies, the largest gravitationally bound systems in the Universe, reside at the nodes of the Cosmic Web \cite{Bond:1996} and are connected by a multitude of filamentary structures. In the outskirts of the clusters, matter and galaxies are funneled towards the centre through filaments. A comprehensive knowledge of the filaments connected to galaxy clusters is essential to understand the influence of the environment on galaxy cluster properties and evolution. One way to quantify the filamentary skeleton around the cluster is through the so-called connectivity $k$ \cite{Codis:2018}, defined as the number of filaments globally connected to the galaxy cluster, estimated at a specific aperture. This project aims to investigate the impact of filaments on the main galaxy cluster properties, in particular in this work we focus on their masses and dynamical state. 
\section{\textrm{\textmd{\trecento}} project}
\label{trecento}
In this work, we analyse the multiple zoom-in regions of \trecento hydrodynamical simulation \cite{Cui:2018}. \trecento project\footnote{\href{https://www.nottingham.ac.uk/~ppzfrp/The300/}{https://the300-project.org}} aims to model 324 regions whose volumes have cubic side lengths of $30 \, h^{-1}\, Mpc$, centered on massive galaxy clusters, with a mass $M_{200} > 6.42 \times 10^{14}  \, h^{-1} \, M_{\odot} $. \trecento regions are re-simulated with higher resolution from the 324 most massive galaxy clusters at $z=0$ of the $1 \, h^{-1} \, Gpc$ Dark Matter-only MDPL2 MultiDark \cite{Klypin:2016} simulation. The cosmological parameters used in the MDPL2 and \trecento simulations are those measured by the \textit{Planck} mission \cite{Planck Collaboration:2016}. The regions have been re-simulated with different baryonic models: \textsc{Gadget-MUSIC} \cite{Sembolini:2013}, \textsc{Gadget-X} \cite{Rasia:2015}, which is used in this work, and more recently \textsc{Gizmo-Simba} \cite{Cui:2022}. For each region, 128 snapshots are available from redshift $z=17$ to $z=0$. The simulated regions were analysed using the \textsc{AHF} halo finder \cite{Knollmann:2011} which self-consistently includes both gas and stars in the halo finding process. The halo finder extracts haloes and estimates their properties, such as the radius $R_\Delta$\footnote{The subscript $\Delta$ indicates the overdensity, i.e. value of the ratio between the density of the cluster at that radius and the critical density of the Universe $\rho_c = 3H^2 / (8 \pi G)$ at the cluster's redshift. }, mass $M_{\Delta}$ and density profile. 

\section{Methods}
\label{Methods}
\subsection{Cosmic Web extraction}
The gas particle distribution of \trecento regions are analysed with DisPerSE \cite{Sousbie:2011}, a topological structure finder, designed to extract the structures of the Cosmic Web.
The finder identifies topologically significant features in the input density field, which is obtained through a Delaunay tesselation in the case of a 2D or 3D discrete distribution. The noise introduced by the finite sampling of the distribution is quantified and reduced with the persistence and topological simplification theories. The persistence parameter, which quantifies the robustness of a topological pair, is defined as the difference of the values of the critical points in the pair and is used to filter low significant filaments. As an output, DisPerSE provides the positions of the extreme points found in the distribution: maxima, minima, saddle points and bifurcation points, where a filament splits in two. The filaments are given as a set of segments connecting a maximum and a saddle point. \\
For our analysis, we first binned the gas particle distribution of each region at redshift $z=0$ in a three-dimensional grid of $30 \, h^{-1} \, Mpc$ per side, and within each region we give a pixel resolution of $150 \, h^{-1} \, kpc$. Then, to avoid sharp variations from one pixel to another, we applied a Gaussian smoothing with a $\sigma$ of 4 pixels. Finally we applied an absolute persistence cut of $0.2$, to focus on significant filaments connecting clusters and haloes. 
The node distribution extracted from DisPerSE was compared with the \textsc{AHF} halo catalogue of each region, to match the DisPerSE maximum points to the simulated haloes and clusters. To avoid a possible low-resolution contamination near the borders, we consider only the haloes inside a sphere of $13  \, h^{-1} \, Mpc$ radius from the centre of each region. The final data set includes $ 3 \times 10^3$ haloes and clusters with a mass range from $10^{13} \leq M_{200} \, h^{-1} \, M_{\odot} \leq 5 \times 10^{15} $.  \\
The gas skeleton extracted from \trecento simulated regions is a good tracer of the overall matter distribution and accretion to galaxy clusters. In particular, throughout the 324 simulated regions there is a good spatial agreement between the gas filaments and both the Dark Matter and mock galaxy filaments, both from 3D and 2D extractions \cite{Kuchner:2020,Kuchner:2021}.
\subsection{Connectivity measurements}
For each halo we estimated the connectivity $k$, which is defined in \cite{Codis:2018} as the number of filaments globally connected to a cluster.
Different definitions are used in literature to estimate this parameter. In this work, we compute the connectivity as the number of filaments crossing a specific spherical surface at a radius $R_\Delta$ from the centre of the halo. In particular, we estimate the connectivity at $R_{200}$ and at $R_{500}$, defined respectively as $k_{200}$ and $k_{500}$. With this definition, we are taking into consideration also the filaments that are coming from substructures and bifurcation points that lie within the sphere, which contribute to the cluster's properties and therefore also to its connectivity.

\section{Results}
\label{Results}
In this work we investigate if the number of filaments connected to a cluster, quantified by the connectivity, is correlated with the main properties of the cluster itself, mainly its mass and dynamical state. 
\subsection{Connectivity and galaxy cluster mass}
\begin{figure}[h]
\centering
\includegraphics[scale=0.8]{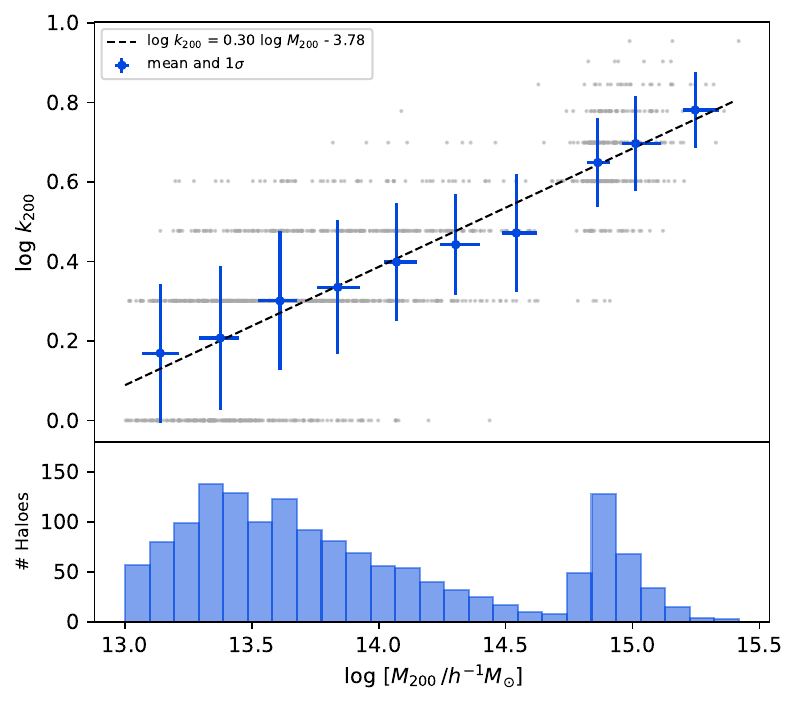}
\caption{The connectivity of haloes and clusters $\log k_{200}$ plotted as a function of the mass $\log \, (M_{200}/h^{-1} M_{\odot})$ (grey points). In the top panel the mean and standard deviation values are also plotted. The bottom panel shows the mass distribution of haloes and clusters of \trecento hydrodynamical simulation at redshift $z=0$ analysed in this work.}
\label{fig:mk scal}      
\end{figure}

The number of filaments connected to a cluster is expected to correlate with the mass of the cluster itself, as many studies show \cite{Aragon-Calvo:2010, Sarron:2019, Darragh Ford:2019,Malavasi:2020}. The expected trend is for the connectivity to increase with the mass of the cluster. \\ 
Taking the advantage to investigate a cluster's sample with the largest mass range, we analyse the connectivity of a set of haloes and clusters extracted from \trecento simulation at redshift $z=0$. Figure~\ref{fig:mk scal} shows the values of the connectivity $k_{200}$ as a function of the clusters mass $M_{200}$. We measured the mean and the standard deviation for each mass bin, chosen by taking into account the overall mass distribution in \trecento simulation at $z=0$, shown in the bottom panel of the figure. We performed a linear fitting, whose parameters are shown in Table \ref{tab:param}, as the following: $ \log k_{200} = A \cdot \log M_{200} + B $.

\begin{figure}[h]
\centering
\includegraphics[scale=0.8]{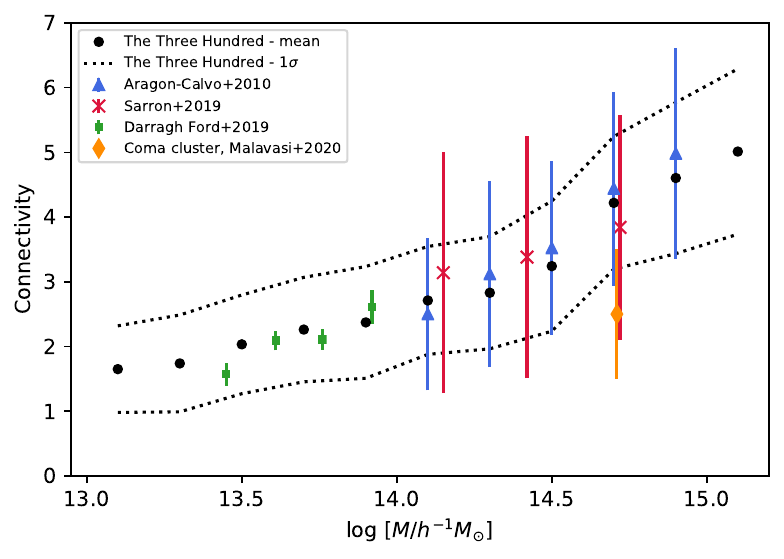}
\caption{Connectivity of \trecento clusters, compared with literature
values. For each work, we plot the mean values of the connectivity. The error bars refer to the standard deviation values for this work, \cite{Aragon-Calvo:2010,Sarron:2019,Malavasi:2020}, while for \cite{Darragh Ford:2019} they represent the errors on the mean values. \label{fig: mk comp}}       
\end{figure}

\begin{table}[h]
    \centering
    \caption{The fitted parameters for the $\log k_{200} - \log M_{200}$ relation}
    \begin{tabular}{cc}
    \hline
    & $\log k_{200} - \log M_{200}$\\
    \hline
    A & 0.298 $\pm$ 0.016  \\
    B & -3.78 $\pm$ 0.23 \\  
    \hline

    \end{tabular}
    \label{tab:param}
\end{table}
We compare our estimates of the connectivity to similar measurements from literature studies, as shown in Figure \ref{fig: mk comp}. We compare our sample both to simulated, as \cite{Aragon-Calvo:2010}, and observed data, such as \cite{Sarron:2019,Darragh Ford:2019,Malavasi:2020}. The trend of the connectivity with the mass is well in agreement, within the errors, with the ones from literature, despite the differences  presented in Table \ref{tab:comparison}, and over a larger mass range. 
More specifically, the connectivity is estimated at a fixed aperture in \cite{Aragon-Calvo:2010, Sarron:2019}, while the connectivity from our work and \cite{Darragh Ford:2019,Malavasi:2020} is estimated at an overdensity radius. Other differences can arise from the Cosmic Web component chosen to extract the filamentary skeletons, such as gas, Dark Matter or galaxies particles and from the different filament finder used in the analyses. We refer to \cite{Aragon-Calvo:2010,Sarron:2019,Darragh Ford:2019, Malavasi:2020} for a more detailed description of the datasets used. 

\begin{table}[h]
    \centering
    \caption{A summary of the parameters of previous studies compared to this work.}
    \begin{tabular}{lllll}
    \hline
    & Data & $k_R$ & M & CW extraction \\
    \hline
    The300 & Hydro simulation & $R_{200}$ & $M_{200}$ & 3D gas particles \\
    AC+10 \cite{Aragon-Calvo:2010} & DM simulation & 3 Mpc & $M_{Vir}$ & 3D DM particles  \\
    S+19 \cite{Sarron:2019} & Observations & 1.5 Mpc & $M_{200}$ & 2D galaxies  \\
    DF+19 \cite{Darragh Ford:2019} & Observations & 1.5 $R_{Vir}$ & $M_{200}$ & 2D galaxies  \\
    M+20 \cite{Malavasi:2020} & Observations (Coma Cl.) & $R_{Vir}$ & $M_{200}$ & 3D z-slice galaxies \\
    \hline

    \end{tabular}

    \label{tab:comparison}

\end{table}

\subsection{Connectivity and galaxy cluster dynamical state} 
\begin{figure}[h]
\centering
\includegraphics[scale=0.45]{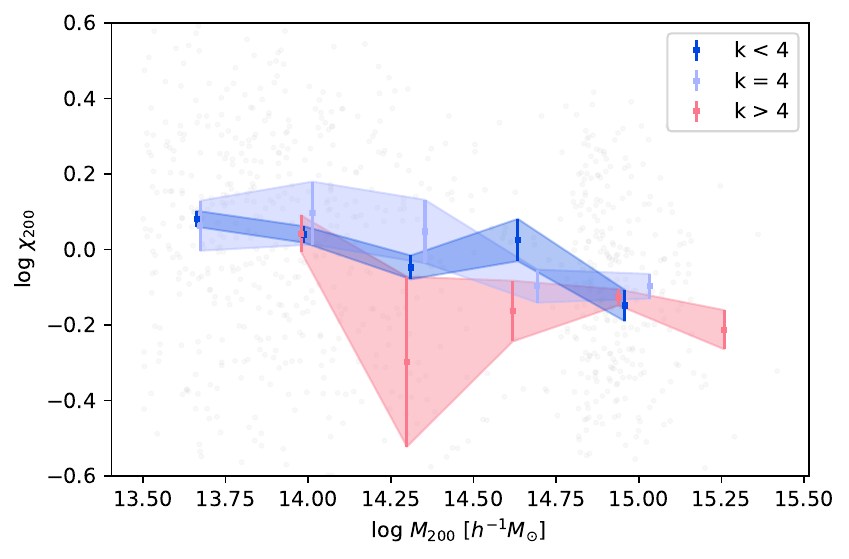}
\includegraphics[scale=0.45]{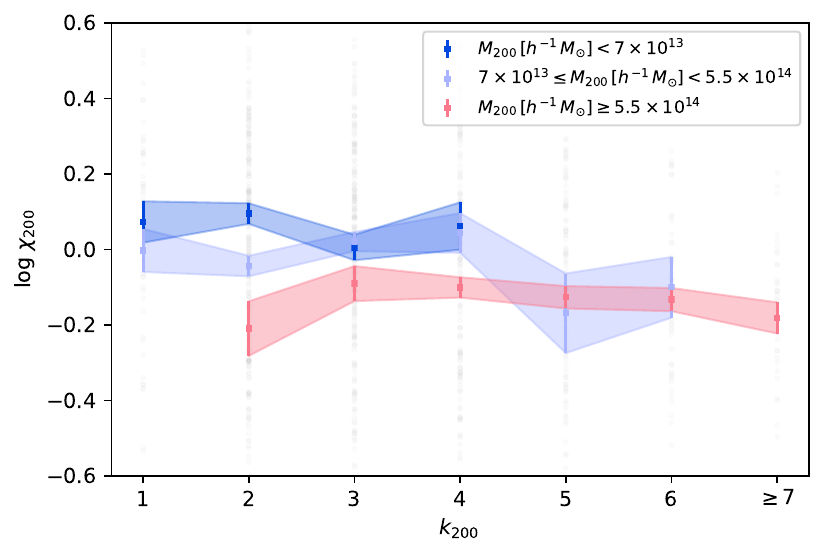}
\caption{\textit{Left panel}: mean degree of relaxation as a function of the mass $M_{200}$ for three connectivity $k_{200}$ bins ($k<4$, $k=4$ and $k>4$). \textit{Right panel}: mean degree of relaxation as a function of the connectivity $k_{200}$ for three mass $M_{200}$ bins. In both panels, the errors bars represent the errors on the mean values.}
\label{fig: dyn state}       
\end{figure}

In this subsection we investigate the correlation between the connectivity and the dynamical state of \trecento clusters. We quantify the dynamical state with the degree of relaxation $\chi$, as defined by \cite{Haggar:2020}:
$$   \chi_{\Delta} =  \left(\frac{(\frac{f_s}{0.1})^2 + ( \frac{\Delta_r}{0.04})^2 + (\frac{|1-\eta|}{0.15})^2}{3}\right)^{-1/2}$$
where $f_s$ is the sub-halo mass fraction, $\Delta_r$ is the centre-of-mass offset and $\eta$ is the virial ratio. The threshold values for these parameters were chosen following \cite{Cui:2017}. A cluster is considered as dynamically relaxed when $\chi \geq 1$. 
To study the effect of connectivity on the dynamical state of clusters, independently on their masses, we divide the data set in three different connectivity and mass sub-samples. Respectively, we consider weakly connected ($k_{200} < 4$), medium connected ($k_{200} = 4$) and highly connected ($k_{200} >4$) clusters along with low mass ($M_{200} < 7 \times 10^{13} \, h^{-1} M_{\odot}$), medium ($7 \times 10^{13} \leq M_{200} \, h^{-1} M_{\odot}< 5.5 \times 10^{14} $) and massive ($M_{200} \geq 5.5 \times 10^{14} \, h^{-1} M_{\odot}$) clusters. \\
In the left and right panels of Figure \ref{fig: dyn state} we display the degree of relaxation as a function of the mass and connectivity, respectively, for three connectivity and mass bins. At a fixed mass, the left panel of Figure \ref{fig: dyn state} shows that there is no evident correlation between connectivity and the degree of relaxation, as the three sub-samples overlap. On the other hand, at fixed connectivity, the right panel of the figure shows a slight correlation between the mass and $\chi$, indicating that less massive haloes are on average more dynamically relaxed. 
This result is in disagreement with that found by \cite{Gouin:2021}, where at fixed mass, weakly connected clusters are on average more relaxed than highly connected clusters. These differences may depend on the different data set and Cosmic Web extraction analysed in their work, to which we refer the reader for more details.

\section{Conclusions}
\label{concl}
In this work we analysed the gas filamentary structures connected to \trecento hydrodynamical simulation clusters and their impact on galaxy cluster properties. We extracted the gas skeletons at $z=0$ with the DisPerSE filament finder in the 324 regions of the simulation and we estimated the connectivity $k_{200}$ of haloes and clusters. \\ 
The main conclusions of this work can be summarized in the following: 
\begin{enumerate}
    \item The connectivity is correlated with the mass of haloes and clusters, with more massive clusters being on average more connected. This result is compatible with previous results from literature, both from simulations and observations;
    \item We do not find a correlation between the connectivity and the dynamical state of clusters, quantified in terms of the degree of relaxation $\chi$. 
\end{enumerate}

%
%

\end{document}